\documentclass[
    aps,
    prl,
    twocolumn,
    superscriptaddress
]{revtex4-2}
\usepackage{amsmath,amssymb,graphicx,xcolor,bm,soul}
\usepackage{hyperref}
\setcitestyle{super,sort&compress}
\hypersetup{colorlinks=true,citecolor={blue},linkcolor={blue},urlcolor={blue}}

\usepackage[normalem]{ulem}
\usepackage{cancel}
\usepackage{tikz}
\usetikzlibrary{arrows.meta}

\usepackage{siunitx}
\usepackage{xcolor}
\definecolor{myred}{HTML}{e76254}
\definecolor{myblue}{HTML}{0094FB}
\definecolor{myblueEQ}{HTML}{198fe2}
\definecolor{myAmethyst}{HTML}{6C2AB8}

\AtBeginDocument{%
  \hypersetup{
    citecolor=myblueEQ,
    linkcolor=myblueEQ,   
    urlcolor=myblueEQ}}

\makeindex

\begin{document}
\date[]{\textbf{Correspondence:} kevin.dini74@gmail.com, simone.zanotti@ntu.edu.sg, timothyliew@ntu.edu.sg}

\title{Spin Simulators based on Exciton-Polariton Waveguide Networks}

\author{Kevin Dini}\thanks{K.D. and S. Z. contributed equally to this work as first authors.}
\affiliation{Division of Physics and Applied Physics, School of Physical and Mathematical Sciences, Nanyang Technological University 637371, Singapore}

\author{Simone Zanotti}
\thanks{K.D. and S. Z. contributed equally to this work as first authors.}
\affiliation{Division of Physics and Applied Physics, School of Physical and Mathematical Sciences, Nanyang Technological University 637371, Singapore}

\author{Timothy C. H. Liew}
\affiliation{Division of Physics and Applied Physics, School of Physical and Mathematical Sciences, Nanyang Technological University 637371, Singapore}
\affiliation{Majulab International Joint Research Unit UMI 3654, CNRS, Université Côte d’Azûr, Sorbonne Université, National University of Singapore, Nanyang Technological University, Singapore}
\affiliation{Centre for Quantum Technologies, Nanyang Technological University Singapore}

\begin{abstract}

Many NP-hard optimization problems can be mapped onto interacting spin systems, whose ground state encodes the optimal solution. Polaritonic implementations based on gain-dissipative dynamics have recently emerged as promising analog computing platforms owing to their intrinsic parallelism and ultrafast operation. However, existing approaches are hindered by two fundamental limitations: maintaining the homogeneous amplitudes required for faithful problem encoding remains challenging in realistic physical systems, and most architectures are restricted to nearest-neighbour interactions, severely limiting the variety of optimization problems that can be  efficiently addressed. We introduce a feasible polaritonic XY simulator based on a reconfigurable waveguide platform that overcomes both limitations. Our approach guarantees amplitude homogeneity while enabling the implementation of arbitrary symmetric coupling matrices, thereby allowing the direct encoding of a wide variety of XY Hamiltonians. These results establish a practical architecture for programmable polaritonic spin simulators and pave the way towards scalable optical hardware for analog optimisation.

\end{abstract}

\maketitle

\subsection*{Introduction}
The field of exciton-polaritonics seeks to use combinations of the coherent, nonlinear, and ultrafast properties of the hybrid optoelectronic system of quantum wells placed in microcavities to develop new device concepts. The field has largely been driven by the idea of a polariton laser, which was recently achieved with electrical-injection at room temperature after a decades-long research effort and a paradigm shift to a waveguided in-plane cavity geometry~\cite{Souissi2026}  rather than a more standard vertical cavity geometry. Aside a source of coherent light being a device in its own right, polariton lasers should also become the power sources of other polaritonic device concepts. Logically, the change to a waveguided geometry requires revisiting concepts originally envisioned with the standard planar microcavity geometry.

Polariton simulators have aimed to use lattices of polariton condensates, arranged in a particular pattern, to represent the ground state of a particular spin graph~\cite{Berloff2017}. In particular, the spatial overlap of neighbouring condensates results in a coupling and under application of a gain (non-resonant excitation) the system lases in the collective state of highest gain. The phases of the polaritons at each condensation site are expected to be mapped to a state of an effective XY Hamiltonian. The aim of the polariton simulator is then to form in a state corresponding to this Hamiltonian’s ground state. This is equivalent to solving an NP-hard problem with vast applications in the optimization of scenarios in logistics, communications, networks, science, and materials design~\cite{Stein2013}.

\begin{figure*}[t]
\centering
\includegraphics[width=\textwidth]{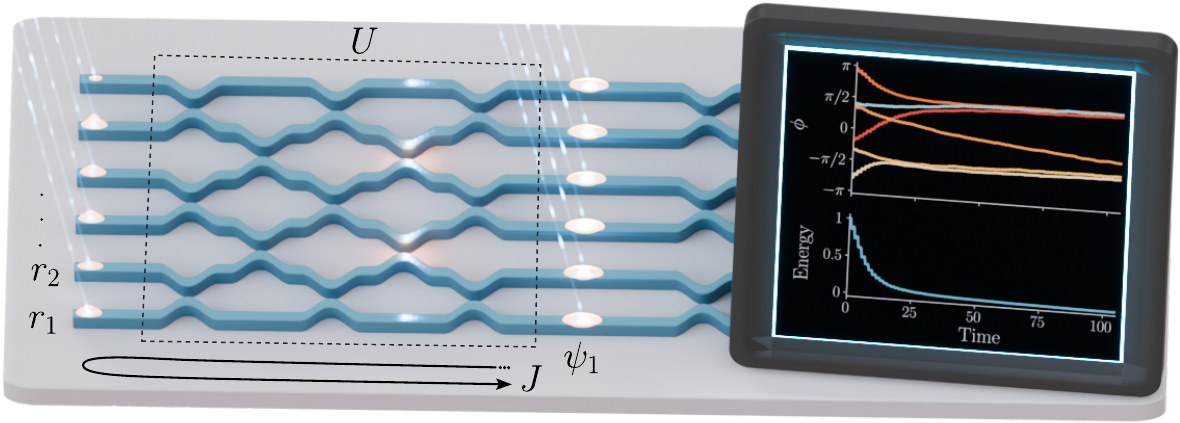}
\caption{{\bf Scheme.}
A set of waveguides are non-resonantly driven at their centre by a main pump modulated with a square-wave temporal profile, which periodically ensures a homogeneous condensate amplitude across the waveguides. The condensate wavefunctions $\psi_i$ at the pump locations encode the spin variables through their phases $\phi_i$. The resulting signals pass through a network of linear beam splitters and phase shifters implementing a unitary transformation $U$. At the opposite ends of the waveguides, non-resonant pumps generate condensates that act as controllable reflectors with reflection coefficients $r_i$. 
This effectively results in a feedback to the driven waveguide fields given by a weighted sum of the other waveguide fields. This sum is associated to an arbitrary all-to-all connected coupling matrix, $J$. 
}
\label{fig:Scheme}
\end{figure*}

Although the aforementioned scheme has been highly successful with experimental demonstrations~\cite{Berloff2017,Luo2020,Harrison2022,Tao2022,Peng2024}, we respect two important limitations. First, as the interaction between condensate centers is linked to their physical separation distance (or control of a barrier between them~\cite{Ohadi2016,Ohadi2016b,Alyatkin2020}), it is not obvious how to engineer arbitrary Hamiltonians. Aside the restriction to near neighbour couplings, the geometry of the graph pattern in which the condensates are excited puts limitations on the combinations of allowed couplings. In theory, one could use external optics to engineer a form of feedback to achieve more arbitrary couplings~\cite{Kalinin2020}, however, this is experimentally challenging. Embedding techniques can be used to expand the range of accessible Hamiltonians, but at the cost of requiring larger graphs~\cite{Harrison2023}.

Second, the mapping of phases to effective spins depends on the phase being the only variable in the system. In fact, a polariton condensate may also vary intensities between different sites to achieve the highest collective gain. This creates a problem of “broken mapping due to amplitude heterogeneity”, which has been identified as a general problem of gain-dissipative solvers~\cite{Kalinin2018}. In theory, a mechanism of local feedback has been proposed to circumvent this issue, resulting in the “Gain-D” algorithm, which represents a state-of-the-art in the field~\cite{Kalinin2018b}. Experimental implementation has not yet been achieved and even in theory the success rate is not 100\%. Nevertheless, the gain-D algorithm is already a powerful algorithmic tool, inspired by but not necessarily needing a physical polariton-based implementation for its use.

Here, we propose a scheme for a polariton simulator based on polariton waveguides. In these structures, light propagates in the plane and can access long propagation lengths due to total internal reflection. It is already experimentally established that propagating polariton signals can be electrically~\cite{Liran2018} or optically~\cite{Suarez2021,Coriolano2025} controlled via local Stark shifts, while waveguides can be coupled to result in effective beam splitters~\cite{Fieramosca2025}. In principle, these ingredients make possible the Reck-Clements scheme~\cite{Reck1994,Clements2016}, where a network of waveguides would allow implementation of an arbitrary unitary operation $U$. The elements of this matrix are ultimately determined by full-field equations, studied in depth in Ref.~\cite{Regemortel2025}. Rather than repeating such analysis here, we take it as previously established that such a matrix exists and is controllable~\cite{Regemortel2026}.

Given this, we consider the situation in Fig.~\ref{fig:Scheme} where a set of polariton waveguides are driven non-resonantly at their mid-points. Following condensation, the resulting signals pass through the network, described by the unitary matrix $U$. Local energy shifts can also be used to generate controlled reflections, described by the (diagonal) reflection matrix $R$. The result is an effective set of feedback terms described by the combination of matrices $U^{-1}RU = J$, where $J$ is a symmetric matrix. Within this scheme $J$ is controllable {\it in situ} if the local phase shifters and reflectivities are controllable (e.g., optically or electrically, with existing technology). Further, $J$ is arbitrary and supports all-to-all coupling.

To preserve the mapping to the XY model, it is important that the polariton intensity remains homogeneous across the waveguides. To achieve this, we propose a pulsed excitation scheme that temporally separates the stabilisation of the condensate amplitudes from the feedback dynamics. As we show below, this periodic restoration of homogeneous amplitudes enables the phase dynamics to implement the desired XY coupling while preventing amplitude heterogeneity from disrupting the mapping.
\subsection*{Results}
{\bf Scheme.---} Our objective is to have a system that is capable of finding the angles $\phi_i$ that minimize an arbitrary XY-Hamiltonian:
\begin{equation}
H=\sum_{ij}J_{ij}\cos(\phi_i-\phi_j)\label{eq:XY}
\end{equation}
where $J_{ij}$ are the elements of a real (and generally dense) symmetric matrix.

Consider a set of non-resonantly driven polariton waveguides each supporting a field $\psi_i(t)$. These fields evolve according to the equations:
\begin{align}
\frac{\partial\psi_i(t)}{\partial t}&=\left(P(t)-1-|\psi_i(t)|^2\right)\psi_i(t)+F_i(t)\notag\\
&\hspace{10mm}+\sigma\eta_i(t)\label{eq:StochasticGP}
\end{align}

Here $P(t)$ is the gain of the waveguide modes, which competes with linear loss and nonlinear losses (which are both normalized to unity). $F_i(t)$ is a feedback field that we will describe later. In the last term, $\eta_i(t)$ is a normalized complex Gaussian white noise, given a strength $\sigma$ to represent amplitude and phase fluctuations in the system. In the above equation, we have neglected the internal dynamics of the gain saturation (i.e., the exciton reservoir dynamics). This is often a good approximation and matches experiments in similar situations~\cite{Peng2024}, while  a full discussion of limitations was presented in Ref.~\cite{Chestnov2024}.

The non-resonant pump $P(t)$ is localised at the centre of each waveguide, as illustrated in Fig.~\ref{fig:Scheme}. Without explicitly modelling their spatial dependence, we assume that each pump generates two polariton fields that travel in equal and opposite directions along the waveguides.
Following the Reck-Clements scheme, it is possible for the waveguides to be arranged with a suitable sequence of (controllable) phase shifters and beam-splitters such that the fields $\psi_i(t)$ effectively undergo a unitary transformation $U$~\cite{Regemortel2025,Regemortel2026}.

After passing through this network, the fields are reflected in a temporally controlled manner. A key feature of our scheme is the temporal separation of the condensation and feedback dynamics, which decouples the stabilization of the condensate amplitudes from the application of the coupling transformation. To this end, the reflection and the pump are temporally gated in a complementary manner. In practice, the reflection coefficients can be implemented through electrical or optical modulation and collected into the diagonal reflection matrix $R$, which is switched on when the pump is off, and vice versa. This modulation naturally defines two distinct intervals: a \textit{feedback period}, during which the reflection is active and the pump is switched off, and a \textit{condensation period}, during which the reflection is switched off and the pump is active. The feedback amplitudes are typically chosen to be smaller than the signal amplitudes, which is favourable for the stability of the feedback dynamics.

During the condensation period, the system is driven by the pump and evolves toward a steady condensate state, which, neglecting noise, is given by:
\begin{equation}\label{eq:cond_sol}
\psi_i=\sqrt{P_0-1}e^{i\phi_i}
\end{equation}
While the intensities of the fields are fixed, the phases $\phi_i$ for the first pulse are random, having been determined by the choice of noise $\eta_i(t)$ during the condensation process.

During the feedback period, the fields propagating in the waveguides are reflected and propagate back through the network, corresponding to the application of the inverse unitary transformation $U^{-1}$. The reflected fields finally return to the centre of the waveguides and are described by $U^{-1}\,R\,U\,\boldsymbol{\psi}(t-\tau)$, where $\tau$ is the full round-trip propagation time through the network. In the absence of the pump, the instantaneous condensate amplitudes $\boldsymbol{\psi}(t)$ decay, while the delayed amplitudes $\boldsymbol{\psi}(t-\tau)$ originate from the preceding condensation period, when their amplitudes were stabilised by the pump and homogeneous across the waveguides.

The overall feedback transformation is therefore given by $U^{-1}RU=J$, where $J$ is chosen to be a real symmetric coupling matrix. Upon returning to the centre of the waveguides, these fields coherently interfere with the local condensate fields, giving rise to the feedback field
\begin{equation}
F_i(t)=-\sum_j J_{ij}\psi_j(t-\tau).
\label{eq:FeedbackFi}
\end{equation}

During the next condensation period, the return of the pump pulses restablishes a uniform intensity, while the phases are otherwise maintained. Our hypothesis is then that, after several pulses, the system will stabilize in a state corresponding to the ground state of an effective XY-Hamiltonian. It should be the state with the smallest loss during the feedback period that is most likely to be maintained in the system. This state minimizes:
\begin{equation}
\sum_{ij} J_{ij} \psi_i^* \psi_j \propto \sum_{ij} J_{ij} \cos\left(\phi_i-\phi_j\right) 
\end{equation}

This proportionality assumes that all the $\psi_i$ have the same intensity (i.e., amplitude homogeneity). This is not strictly true throughout the whole sequence, but at least true at the onset of the feedback periods when $P(t)$ is switched off. There may be some variations in $|\psi_i(t)|^2$ for different waveguides, however, uniform intensities will be re-established when $P(t)$ is switched on again.

\begin{figure*}[t]
\centering
\includegraphics[width=\textwidth]{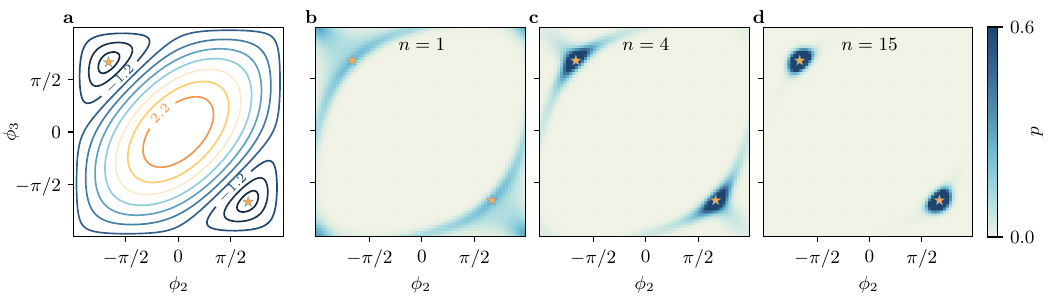}
\caption{{\bf Antiferromagnetic triangle phase distribution.} \textbf{a} Energy landscape of the antiferromagnetic triangular system described by the XY Hamiltonian in Eq.~\eqref{eq:XYTriangle_main}, with $\phi_1=0$. The two degenerate minima are indicated by the stars, and representative contour lines are labelled by their energy values. \textbf{b}--\textbf{d} Phase probability distribution calculated using the recursive protocol of Eq.~\eqref{eq:ProbabilityPhi} for the same triangular system. The repeated feedback progressively concentrates the probability around the two degenerate ground states. }
\label{fig:prob}
\end{figure*}

{\bf Fokker-Planck Treatment.---} To better interpret the proposed scheme, we first present a simplified semi-analytical treatment of the probability distributions in phase space, assuming that a steady state is reached during each condensation  period of the protocol. This assumption is relaxed later in the full numerical analysis, while a complete derivation of the mathematical formalism is provided in the Supplementary Material (SM). We label successive condensation periods by the index $n$.

To introduce the formalism, we consider the feedback step following the $n$th condensation period, which determines the phase distribution of the subsequent $(n+1)$th condensation period. Since there is no gain saturation ($P=0$), the nonlinear term can be neglected, yielding
\begin{equation}
\frac{\partial\psi_i}{\partial t}=(-\psi_i+F_i)+\sigma \eta_i,
\label{eq:Langevin_main}
\end{equation}
where $F_i=-\sum_j J'_{ij}e^{i\phi_j^{(n)}}$ is a complex constant determined by the phase configuration of the $n$th condensation period. Combining the homogeneous steady-state solution in Eq.~\eqref{eq:cond_sol} with the feedback field in Eq.~\eqref{eq:FeedbackFi}, we define $J'_{ij}=\sqrt{P_0-1}\,J_{ij}$ and $J_{ij}\in\mathbb{R}$; the generalisation to complex couplings is presented in the SM.

Using It\=o calculus, Eq.~\eqref{eq:Langevin_main} can be recast as a Fokker--Planck equation (see SM for the derivation). This reformulation allows us to determine directly the steady-state probability density. The probability density of the $i$th waveguide being in the state $\psi_i$ conditioned on the feedback field $F_i$ is
\begin{equation}
p_i(\psi_i|F_i)=\frac{2}{\pi\sigma^2}e^{-2|F_i-\psi_i|^2/\sigma^2}.
\label{eq:GaussianP_main}
\end{equation}
Since the solution depends only on the ratio between the feedback field and the noise strength, it is convenient to introduce the reduced feedback $
\tilde{F}_i=F_i/\sigma$. Moreover, because the XY Hamiltonian is encoded in the phases of $\psi_i$, it is useful to rewrite Eq.~\eqref{eq:GaussianP_main} in terms of the phase probability,
\begin{align}
p_i(\phi_i|\tilde{F}_i)
&=
\frac{e^{-2|\tilde{F}_i|^2}}{2\pi}
\left\{
1+\sqrt{\pi}\,e^{\tilde{\zeta}_i^2}\,
\tilde{\zeta}_i
\left[
1+\mathrm{Erf}(\tilde{\zeta}_i)
\right]
\right\},
\label{eq:Pphii_main}
\end{align}
where $\mathrm{Erf}$ is the error function, and
\begin{equation}\label{eq:zeta_main}
\tilde{\zeta}_i=\sqrt{2}\,\mathrm{Re}\left\{\tilde{F}_i e^{-i\phi_i}\right\}=-\frac{\sqrt{2}}{\sigma}\,\sum_j J'_{ij} \,\cos(\phi_i-\phi_j^{(n)}).
\end{equation}
$\tilde{\zeta}_i$ is proportional to the opposite of the XY interaction energy associated with site $i$.

Given the feedbacks $\tilde{F}_i$, the Langevin equations \eqref{eq:Langevin_main} evolve independently. Consequently, the phases $\phi_i$ are conditionally independent, and the joint conditional probability factorises as:
\begin{equation}
p(\boldsymbol{\phi}|\boldsymbol{\tilde{F}})=\prod_i p_i(\phi_i|\tilde{F}_i).
\end{equation}

The conditional probability can therefore be written equivalently as
$
p_i(\phi_i|F_i)=p_i(\phi_i|\boldsymbol{\phi}^{(n)}),
$
where $\boldsymbol{\phi}^{(n)}=(\phi_1^{(n)},\ldots,\phi_N^{(n)})$, with $N$ denoting the number of waveguides.
The unconditional probability distribution is then obtained by integrating over the probability distribution of the previous phase configuration. Since each condensation cycle generates the feedback field for the next one, this naturally leads to a recursive evolution of the phase distribution. Introducing $p^n(\boldsymbol{\phi})$ as the probability distribution after the $n$th feedback event, we obtain
\begin{equation}
p^{n+1}(\boldsymbol{\phi})=\int \left[\prod_i p_i(\phi_i|\boldsymbol{\phi}^{n}) \right]p^n(\boldsymbol{\phi}^{n})\,d\boldsymbol{\phi}^{n}.
\label{eq:ProbabilityPhi}
\end{equation}
Equations \eqref{eq:Pphii_main}, \eqref{eq:zeta_main}, and \eqref{eq:ProbabilityPhi} show that the most likely state is the one that maximises $\sum_i\zeta_i$, or equivalently, minimizes the XY Hamiltonian energy. 

However, after a single feedback step the probability distribution remains broadly distributed over phase space. To determine its long-term evolution, we initialise the phase distribution uniformly and iteratively apply Eq.~\eqref{eq:ProbabilityPhi}. The repeated feedback progressively biases the distribution towards configurations with larger values of $\sum_i\tilde{\zeta}_i$, ultimately converging to the ground state of the XY Hamiltonian.

{\bf Antiferromagnetic Triangle.---} As a preliminary demonstration, let us consider the simple example of the XY Hamiltonian for three spins:
\begin{equation}
\mathcal{H}=J\left(\cos(\phi_1-\phi_2)+\cos(\phi_2-\phi_3)+\cos(\phi_1-\phi_3)\right)\label{eq:XYTriangle_main}
\end{equation}

\begin{figure*}[t!]
\centering
\includegraphics[width=0.49\textwidth]{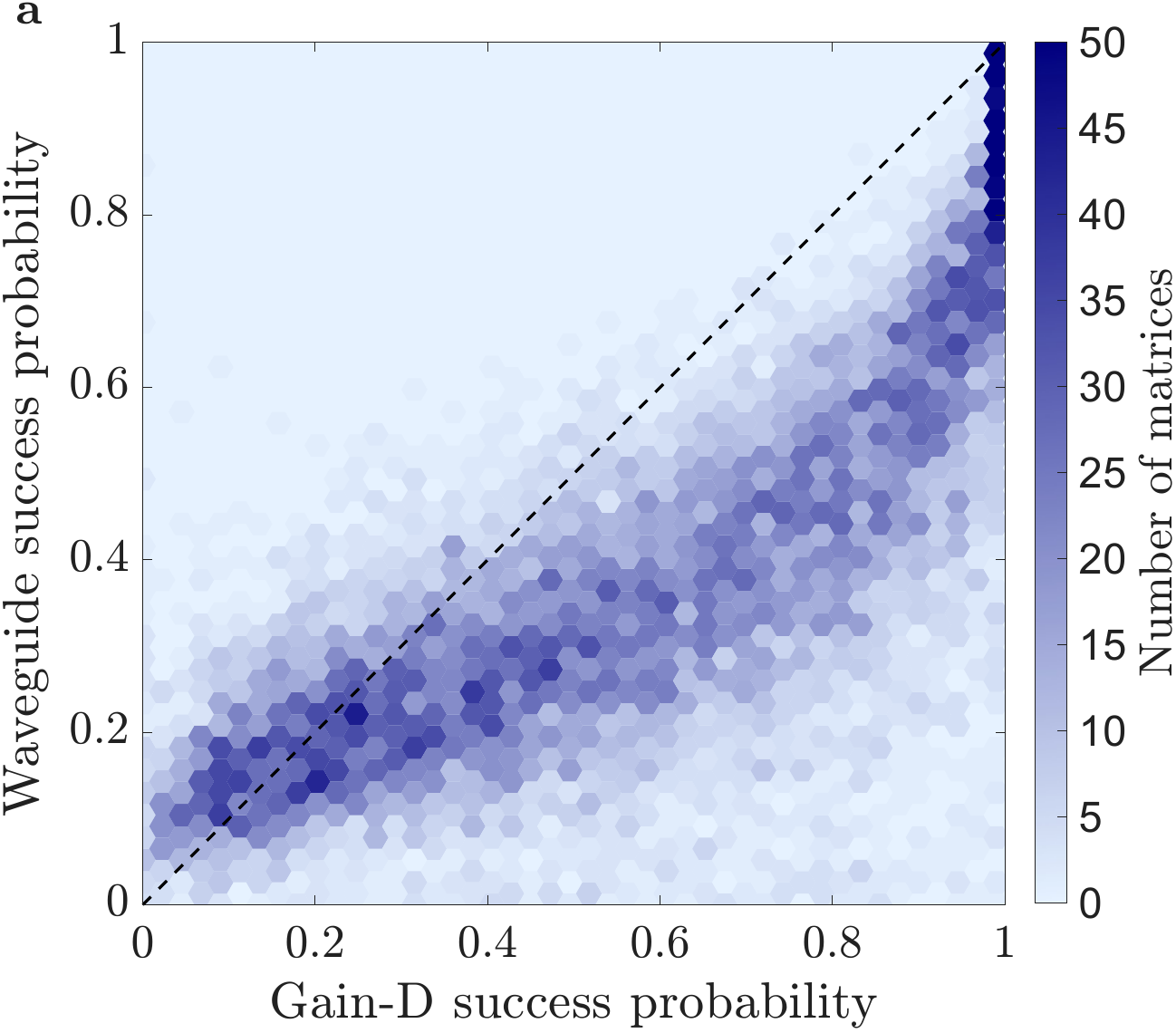}
\hfill
\includegraphics[width=0.49\textwidth]{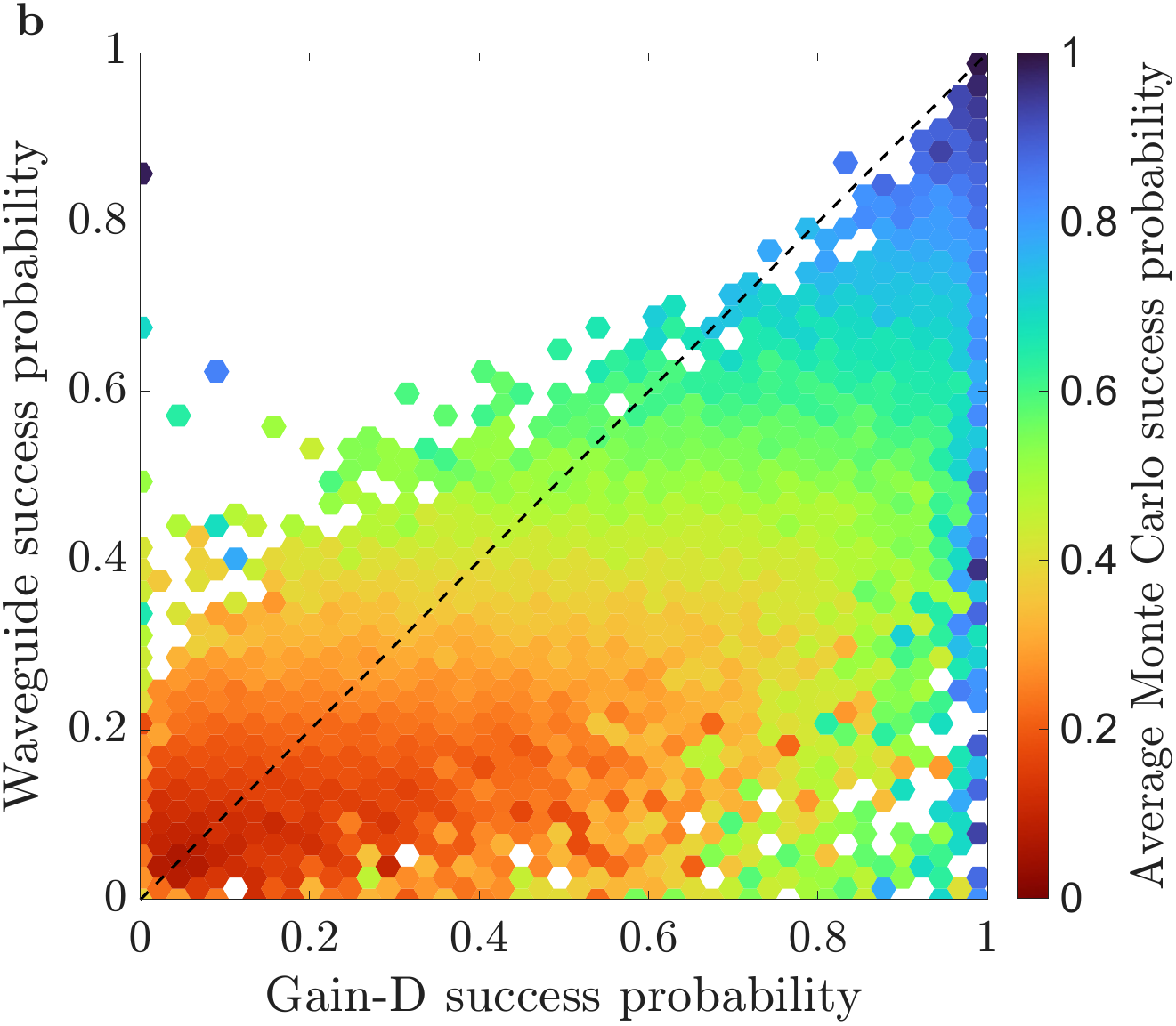}
\caption{\textbf{Success-rate comparison between Gain-D and waveguide solvers.}
 (\textbf{a}) Waveguide (WG) success probability plotted against Gain-D (GD) success probability for \num{12205} randomly generated XY Hamiltonian instances. A realization is considered successful when its relative energy error is below the prescribed numerical tolerance. The color scale indicates the number of matrices in each bin. The dashed diagonal marks equal performance of the two methods. (\textbf{b}) Same comparison, with each matrix colored according to its Monte Carlo (MC) success probability, providing an independent measure of instance hardness. The distribution shows that the relative performance of WG and GD depends strongly on the hardness of the underlying XY instance.}
\label{fig:successRateComparison}
\end{figure*}

As the energy should be invariant with respect to a global phase shift we set $\phi_1=0$ with no loss of generality. The energy landscape is shown in Fig.~\ref{fig:prob}a. There are two degenerate energy minima at $(\phi_2,\phi_3)=(\pm2\pi/3,\mp2\pi/3)$ with energy $-3J/2$. It is instructive to consider the diagonalization~\cite{Cao2025} of the coupling Hamiltonian:
\begin{equation}
\label{eq:coupling_matrix}
\mathcal{H}
=
\begin{pmatrix}
0 & J & J\\
J & 0 & J\\
J & J & 0
\end{pmatrix}
\end{equation}
which yields eigenvectors $(1,1,1)/\sqrt{3}$, $(1,-1,0)/\sqrt{2}$, and $(1,0,-1)/\sqrt{2}$. As the eigenvectors have phases corresponding to $(0,\pi)$, they can not possibly match the ideal ground state solution. This is because the lowest eigenvector not only adjusts its phases to achieve the lowest energy state of the coupling matrix but also its amplitudes. The solution to an XY Hamiltonian corresponds to a constrained optimization problem, where only phases (or spins) are allowed to rotate. This is why gain-dissipative solvers require an additional density adjustment mechanism to reach the correct ground state. Let us also note that it is insufficient to simply reduce the coupling strength $J$, as this will not actually change the eigenvectors of the coupling matrix.

In Fig.\ref{fig:prob}b--d we show the result of iterating Eq.~\ref{eq:ProbabilityPhi} over multiple pulses. We find that the probability distribution gradually converges onto the two ground state solutions, apart from a small spread that remains due to the presence of noise. To attain this result, we found it necessary to introduce self-interaction terms, equivalent to diagonal terms in the coupling matrix. We stress that this does not add anything to the complexity of our proposed scheme and the reason for these terms is discussed in detail in the SM.

{\bf Numerical Analysis.---} The aforementioned Fokker-Planck treatment was intended to elucidate the mechanism of our scheme, however, the evaluation of its performance requires numerical simulations of Eqs.~\ref{eq:StochasticGP} and~\ref{eq:FeedbackFi} with larger system sizes and comparison to established benchmarks. Henceforth, we will refer to our architecture as the waveguide (WG) method. As a standard benchmark, we will compare the results with a Monte-Carlo supplemented with the Nelder-Mead simplex method (MC+). This combines efficient relaxation to local minima with Monte-Carlo global search. Hopping between basins is possible, but not guaranteed. As the best-known method in exciton-polaritonics, we also compare to the original gain-dissipative architecture (GD0) introduced in~\cite{Berloff2017}. As the state-of-the-art in the field, we take the theoretical gain-dissipative method supplemented with density adjustment or gain-D method (GD)~\cite{Kalinin2018b}. 

Our aim is to achieve a solving efficiency comparable to that of GD, while removing the need for active feedback adjustment, which can be experimentally challenging to implement. We assume that all {the aforementioned} methods can realize all-to-all coupling. This allows us to directly compare the methods at the theoretical level, putting aside differences in experimental difficulty of actually implementing all-to-all coupling in a physical realization.

As a standard benchmarking task, we consider real symmetric $50\times 50$ XY coupling matrices, with random coupling strengths drawn from a Gaussian distribution centred at zero, such that both ferromagnetic and antiferromagnetic interactions are present. Strictly speaking, not all matrices generated with this procedure correspond to NP-hard problems. The failure rate of an MC+ type algorithm {is commonly used as} a measure of {problem} ``hardness''\cite{Kim2024,Kim2026,Kalinin2018b}. However, despite the clear correlation between problem hardness and the success probabilities of WG and GD, we find that individual matrix realizations can be harder or easier for different methods.

\begin{figure*}[t!]
\includegraphics[width=1\textwidth]{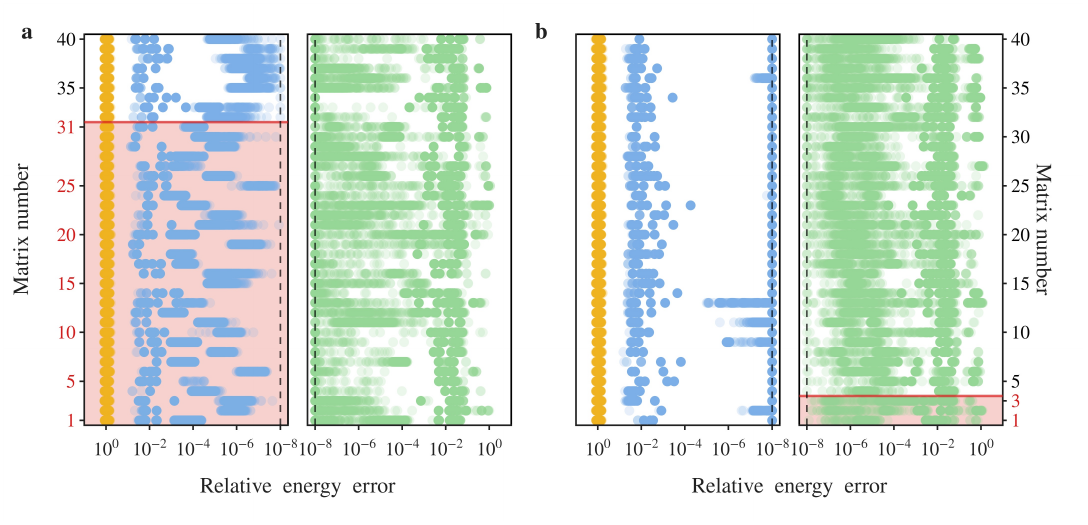}
\caption{{\bf Energy comparison for hard matrices.}
Energy comparison for hard XY instances. From a library of \num{12205}
randomly generated XY matrices, we select the 40 hardest instances for
GD (\textbf{a}) and WG (\textbf{b}), according to their respective success
probabilities. For each selected matrix, GD0 (orange), GD (blue), and WG
(green) are evaluated over 1000 independent noise realizations. Each point
shows the relative energy error with respect to the reference ground-state
energy.}
\label{fig:EnergyComparison}
\end{figure*}

For a statistically robust comparison, we generated \num{12205} matrices. For each matrix, we estimated the ground-state energy using the MC+ algorithm with a large number of samples \num{5000}, ensuring a high confidence in the identified ground state. 
{Using this reference energy, we define the success rate of a method as the fraction of noise realizations whose final energy satisfies a relative error of at most $10^{-8}$ with respect to the reference ground-state energy. We use the success rate as a measure of instance hardness, with lower values corresponding to harder matrices. A zero success rate therefore corresponds to a failure to solve the instance across all noise realizations. }

As a first overview of the general behaviour of GD and WG, Fig.~\ref{fig:successRateComparison}a shows the joint distribution of their success probabilities over all \num{12205} matrices. Since all methods are noise dependent, each success probability is calculated using 1000 independent noise realizations.
The upper-right and lower-left regions correspond to instances that are easy and hard for both methods, respectively, while the dashed diagonal indicates equal success probabilities. Overall, the distribution is predominantly located below the diagonal, indicating a higher success probability for GD for a fixed number of noise realisations. Interestingly, however, within the lower-left region, corresponding to the hardest instances for both methods, a substantial fraction of the distribution lies above the diagonal. This indicates that WG tends to achieve a higher success probability precisely for instances that are difficult for both methods.

To provide an independent measure of instance hardness, Fig.~\ref{fig:successRateComparison}b shows the same success-probability distribution, with each matrix colour-coded according to its hardness for the reference MC+ method. A clear correlation emerges: matrices that are harder for the reference MC+ method also tend to exhibit lower success probabilities for both GD and WG. This correlation indicates that the three methods share, to a significant extent, a common notion of instance hardness.

Having established the general behaviour of the different methods, we next examine the hardest instances, where the difficulty of the underlying optimisation problem is most pronounced. Figure~\ref{fig:EnergyComparison} shows the resulting relative energy errors for the WG (green), GD0 (orange), and GD (blue) methods for the 40 hardest matrices for GD in \textbf{a} and WG in \textbf{b}. 
The hyper-parameters for GD were taken from~\cite{Kalinin2018b} and were found to be near-optimal in our tests,  while those for WG were optimised following the procedure described in the SI.

For every instance considered, the performance of GD is markedly better than that of GD0, with the latter failing to reach the ground state in any of the calculations. This stark difference highlights the need of density adjustment for these architectures, whereas our approach requires no such adjustment.

Across the \num{12205} sampled matrices, the GD method failed for 31 instances, whereas our WG method failed for only 3, as indicated by the red-shaded regions in Fig.~\ref{fig:EnergyComparison}. Interestingly, none of the hardest matrices for GD in Fig.~\ref{fig:EnergyComparison}a belong to the hardest set for WG in Fig.~\ref{fig:EnergyComparison}b . This suggests a degree of complementarity between the two methods: an instance that is particularly challenging for one method can be more readily solved by the other, although it is not known a priori which method will perform better for a given instance.

\begin{figure}[t!]
\includegraphics[width=0.5\textwidth]{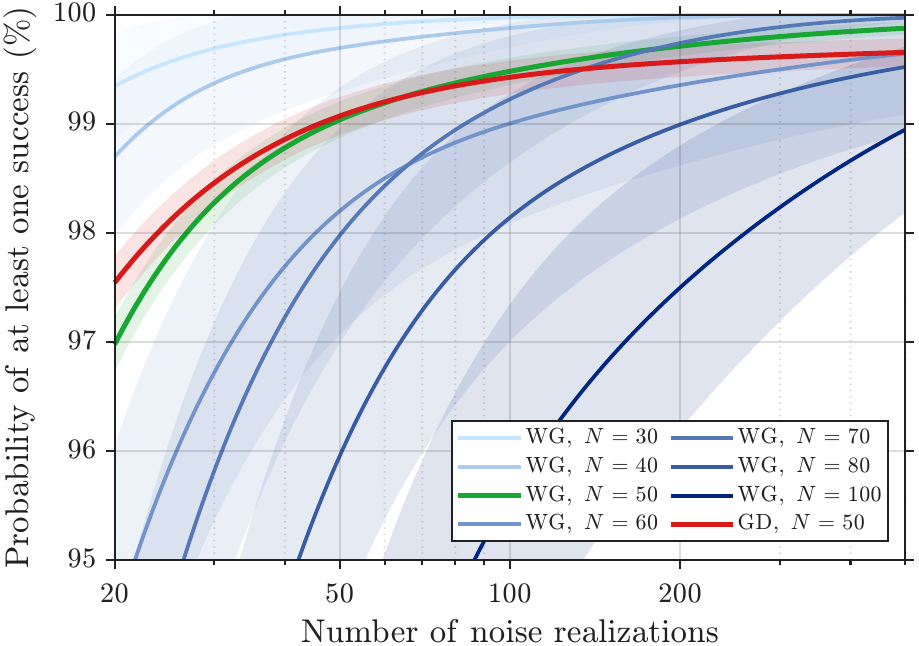}
 \caption{
    Probability of obtaining at least one successful realization as a function of the number of noise realizations. For each matrix, the cumulative success probability is calculated from the single-realization success probability $p_i$ and then averaged over the matrix ensemble (See Supplemenrary materials for more details). Blue curves show the waveguide (WG) results for matrix sizes $N=30$--$100$, with $N=50$ highlighted in green. The Gain-D (GD) result for $N=50$ is shown in red; the WG and GD $N=50$ curves are evaluated over the same set of \num{12205} matrices. Shaded regions indicate 99\% confidence intervals of the ensemble mean. The same data on a wider scale is available in the SI.}
\label{fig:Scalability}
\end{figure}

\subsection*{Discussion}

A critical consideration for any analogue computing platform is its scalability to larger problem sizes. Our waveguide architecture can be directly extended to larger coupling matrices without requiring modifications to the underlying scheme. To assess its performance upon scaling, Fig.~\ref{fig:Scalability} shows the probability of obtaining at least one successful solution after a given number of independent noise realisations. In an experimental implementation, these realisations would correspond to consecutive repetitions of the solving protocol, separated by a short time interval. The calculation of this probability is detailed in the SI.

We perform this numerical analysis using our WG method for increasing problem dimensions, ranging from $20\times20$ to $100\times100$, shown in blue, with the $50\times50$ case highlighted in green. For comparison, we also consider the GD method for $50\times50$ matrices shown in red, for which optimised hyperparameters have previously been reported~\cite{Kalinin2018b}.

Extending this numerical analysis to substantially larger problem sizes while retaining sufficient statistics becomes increasingly computationally demanding. This limitation further motivates the physical implementation of the proposed analogue solver. Ultimately, such an implementation would provide a route towards exploring regimes that become increasingly difficult to access through extensive numerical sampling, which is one of the central motivations of the present proposal. Here we consider coherent Ising machines inspirational, having scaled from an initial experimental demonstration with four coupled nodes~\cite{Marandi2014} to $10^5  $~\cite{Honjo2021} in under a decade. Even though existing realizations of polariton simulators are not at this scale, it was pointed out that a small-sized system could be used in a modular approach as part of a larger network~\cite{Stroev2021}. While spin glass and other NP-hard problems can typically be mapped into one another~\cite{Stein2013}, native encodings of XY Hamiltonians have advantages for specific tasks such as image segmentation~\cite{Harrison2022} and phase retrieval~\cite{Wang2026}. In the SM, we discuss how our scheme can be generalized to the complex XY model (supporting Dzyaloshinskij–Moriya terms), while we also speculate that extension to a vector XY should also be possible, following the exploitation of the spin degree-of-freedom~\cite{Kudlis2025}. Overall, we have proposed the present WG method as a logical step given the recent development of polariton laser sources in waveguide geometry~\cite{Souissi2026}, while admitting that polariton waveguide technology is still too much in its infancy to define its limits.

During the preparation of this manuscript, we became aware of a related work \cite{Yu2026}, which independently identifies the limitations of standard XY simulators in phase encoding under conventional nonlinearity.

{\bf Acknowledgements.---} We were supported by the National Research Foundation (Singapore) grant N-GAP (NRF2023-ITC004-001). TCHL also acknowledges support from the Centre for Quantum Technologies (Singapore) Fellowship. We thank W. Verstraelen and J.Z. Perez for discussions.

{\bf Authors Contributions.---} K.D. and S.Z. contributed equally to this work. All authors contributed to the conceptualization of the study, development of the theoretical framework, numerical simulations, analysis and interpretation of the results, and preparation of the manuscript. All authors reviewed and approved the final version.

{\bf Data Availability.---}
The data that support the plots in this paper are available from the corresponding authors. 

{\bf Conflict of Interest.---}
The authors declare no competing interests.

\onecolumngrid

\renewcommand{\theequation}{S\arabic{equation}}
\setcounter{equation}{0}
\renewcommand{\thesection}{S\arabic{section}}
\setcounter{section}{0}


\begin{center}
	\textbf{\large --- SUPPLEMENTAL MATERIAL ---}\\[0.8em]
    {\large \textit{Spin Simulators based on Exciton-Polariton Waveguide Networks }}
\end{center}

\section{ Fokker--Planck equation from Langevin equation}
In this section, we derive the Fokker--Planck equation corresponding to the Langevin equation (Eq.~\eqref{eq:Langevin_main} of the main text) using It\=o calculus. We then obtain the probability distribution with respect to the wavefunction phases (Eq.~\eqref{eq:Pphii_main} of the main text).

We begin from the Langevin equation
\begin{equation}
\frac{\partial\psi_i}{\partial t}=(-\psi_i+F_i)+\sigma \eta_i,
\label{eq:Langevin}
\end{equation}
and we rewrite it in terms of differential elements (we will drop the $i$ subscript for brevity; noting that all the equations of this section hold independently for each value of $i$):
\begin{align}
d\psi&=(-\psi+F)dt+\sigma dW\label{eq:dpsi}\\
d\psi^*&=(-\psi^*+F^*)dt+\sigma dW^*\label{eq:dpsiconj}
\end{align}
where
\begin{align}
\langle dW\rangle &=0\\
\langle dWdW\rangle &=0\\
\langle dWdW^*\rangle &=dt
\end{align}
Now, let us consider the observable quantity $\xi(\psi,\psi^*)$, which is some (smooth) function of $\psi$ and $\psi^*$. Its average value is given by:
\begin{equation}
\langle \xi\rangle=\int \xi(\psi,\psi^*)\, p(\psi,\psi^*,t)d\psi d\psi^*\label{eq:Pdistribution}
\end{equation}
where we have introduced the (implicitly normalized) probability $p(\psi,\psi^*,t)$ of having the state $\psi$ at a given time $t$. As $\psi$ is complex, we denote here the probability distribution as depending on two variables. Let us now consider the change of $\xi$:
\begin{align}
d\xi&=\frac{\partial\xi}{\partial\psi}d\psi+\frac{\partial\xi}{\partial\psi^*}d\psi^*+\frac{1}{2}\frac{\partial^2\xi}{\partial\psi^2}(d\psi)^2+\frac{\partial^2\xi}{\partial\psi\partial\psi^*}d\psi d\psi^*\notag+\frac{1}{2}\frac{\partial^2\xi}{\partial\psi^{*2}}(d\psi^*)^2
\end{align}
Using Eqs.~\ref{eq:dpsi} and~\ref{eq:dpsiconj}, together with the definitions of averages of $dW$, $dWdW$, and $dWdW^*$, we have:
\begin{equation}
\langle d\xi\rangle=\frac{\partial\xi}{\partial\psi}\langle F-\psi\rangle dt+\frac{\partial\xi}{\partial\psi^*}\langle F^*-\psi^*\rangle dt+\frac{\partial^2\xi}{\partial\psi\partial\psi^*}\sigma^2dt
\end{equation}
where we neglected terms second order in $dt$. The differential rate of change of $\langle \xi\rangle$ is then:
\begin{equation}
\frac{d}{dt}\langle\xi\rangle=\frac{\partial\xi}{\partial\psi}\langle F-\psi\rangle+\frac{\partial\xi}{\partial\psi^*}\langle F-\psi\rangle^*+\frac{\partial^2\xi}{\partial\psi\partial\psi^*}\sigma^2
\end{equation}
which can be re-written in terms of the probability distribution using Eq.~\ref{eq:Pdistribution}:
\begin{align}
&\frac{d}{dt}\int \xi(\psi,\psi^*) p(\psi,\psi^*,t)d\psi d\psi^*\notag\\&\hspace{5mm}=\int \left[\frac{\partial\xi}{\partial\psi}(F-\psi)+\frac{\partial\xi}{\partial\psi^*}(F-\psi)^*+\frac{\partial^2\xi}{\partial\psi\partial\psi^*}\sigma^2\right]p(\psi,\psi^*,t)\,d\psi d\psi^*\notag\\
&\hspace{5mm}=-\int \xi(\psi,\psi^*)\frac{\partial}{\partial\psi}\left[(F-\psi)p(\psi,\psi^*,t)\right]d\psi d\psi^*\notag\\
&\hspace{10mm}-\int \xi(\psi,\psi^*)\frac{\partial}{\partial\psi^*}\left[(F-\psi)^*p(\psi,\psi^*,t)\right]d\psi d\psi^*\notag\\
&\hspace{10mm}+\int \xi(\psi,\psi^*)\frac{\partial^2}{\partial\psi\partial\psi^*}\left[\sigma^2p(\psi,\psi^*,t)\right]d\psi d\psi^*
\end{align}
where in the last line we used integration by parts (the boundary terms vanish as we assume that the probability $p(\psi,\psi^*,t)$ vanishes as $\psi$ or $\psi^*$ go to $\pm\infty$). We now see that $\xi(\psi,\psi^*)$ appears as a common factor in the integrand. As the above equation must hold for any function $\xi(\psi,\psi^*)$, we must have:
\begin{align}\label{eq:FokkPl}
\frac{d}{dt}p(\psi,\psi^*,t)&=-\frac{\partial}{\partial\psi}\left[(F-\psi)p(\psi,\psi^*,t)\right]\notag\\
&\hspace{5mm}-\frac{\partial}{\partial\psi^*}\left[(F-\psi)^*p(\psi,\psi^*,t)\right]\notag\\
&\hspace{5mm}+\frac{\partial^2}{\partial\psi\partial\psi^*}\left[\sigma^2p(\psi,\psi^*,t)\right].
\end{align}

Equation~\eqref{eq:FokkPl} is a Fokker--Planck equation whose steady-state solution is
\begin{equation}
p_i(\psi_i|F_i)=\frac{2}{\pi\sigma^2}e^{-2|F_i-\psi_i|^2/\sigma^2}.
\label{eq:GaussianP}
\end{equation}

\subsection{Phase probability distribution}
The probability distribution in Eq.~\eqref{eq:GaussianP} written in terms of the phase in the $i$th waveguide is obtained by expressing $\psi_i=|\psi_i|e^{i\phi_i}$ and integrating over the wavefunction amplitude $|\psi_i|$:
\begin{align}
&p_i(\phi_i|F_i)=\int_0^\infty p_i(\psi_i|F_i)|\psi_i|d|\psi_i|\notag\\
&\hspace{5mm}=\frac{2}{\pi\sigma^2}e^{-2|F_i|^2/\sigma^2}\int_0^\infty e^{-2|\psi_i|(|\psi_i|-\zeta_i)/\sigma^2}\psi_i|d|\psi_i|\notag\\
&\hspace{5mm}=\frac{2}{\pi\sigma^2}e^{-2|F_i|^2/\sigma^2}\notag\\
&\hspace{10mm}\left(\frac{\sigma^2}{4}+e^{\zeta_i^2/(2\sigma^2)}\frac{\sqrt{2\pi}}{8}\sigma\zeta_i\left(1+\mathrm{Erf}\left(\frac{\zeta_i}{\sqrt{2}\sigma}\right)\right)\right)\label{eq:Pphii}
\end{align}
where $\mathrm{Erf}$ denotes the error function and we defined:
\begin{equation}
\label{eq:zeta}
\begin{split}
\zeta_i
&=-
2\mathrm{Re}\left\{F_ie^{-i\phi_i}\right\}\\
&=
-2\sum_j\mathrm{Re}\{J'_{ij}\}\cos(\phi_i-\phi_j')
\end{split}
\end{equation}
Introducing the reduced feedback amplitudes $\tilde{F}_i=F_i/\sigma$ and $\tilde{\zeta}_i=\zeta_i/(\sqrt{2}\sigma)$, Eqs.~\eqref{eq:Pphii} and~\eqref{eq:zeta} become Eqs.~\eqref{eq:Pphii_main} and~\eqref{eq:zeta_main} of the main text, respectively.

\section{Chiral antiferromagnetic triangle}

In the main text, we considered the archetypal antiferromagnetic triangle described by the Hamiltonian
\begin{equation}
\mathcal{H}=J\left(\cos(\phi_1-\phi_2)+\cos(\phi_2-\phi_3)+\cos(\phi_1-\phi_3)\right).
\label{eq:XYTriangle}
\end{equation}

Here, we consider a more general example to demonstrate that the proposed method remains robust in the presence of additional interactions. Specifically, we consider the generalized XY Hamiltonian
\begin{align}
H&=J\left[\cos(\phi_1-\phi_2)+\cos(\phi_2-\phi_3)+\cos(\phi_1-\phi_3)\right]\notag\\
&+\epsilon J\left[\sin(\phi_1-\phi_2)+\sin(\phi_2-\phi_3)+\sin(\phi_1-\phi_3)\right],
\label{eq:XYTriangleChiral}
\end{align}
which is commonly referred to as the chiral XY model or the XY model with Dzyaloshinskii-Moriya interactions. The additional sine terms, absent from Eq.~\eqref{eq:XYTriangle}, lift the degeneracy of the ground state. For $\epsilon\ll1$, however, they only weakly perturb the phase-space landscape, leaving the sizes of the basins of attraction essentially unchanged. Consequently, the success rate for many solvers (e.g., Monte-Carlo with gradient descent) drops to just above $50\%$, as there are similar chances for falling into the local rather than the global energy minimum.

For our scheme, the additional sine dependent terms are readily realized by introducing an additional phase delay for the waveguide network to the left or right of the local gain pulses. This effectively multiplies the corresponding coupling matrix by the factor $i$. Note that as the left and right-hand sides of the system can in principle realize independent matrix, the overall arbitrary complex coupling becomes possible.

\section{Significance of diagonal terms}

In our previous demonstration, we have added diagonal couplings into the coupling matrix $J$, equivalent to self-couplings in the XY Hamiltonian. These couplings do not change the ground state solution of the XY Hamiltonian apart from an overall energy shift. They do not cost anything in our scheme, as they only require a redefinition of the $J$ coupling matrix to be implemented. However, they can play a relevant role in specific situations.

In the case of the antiferromagnetic triangle, without the diagonal couplings, we would have:
\begin{equation}
F_{2,3}(\phi_2',\phi_3')=F(\phi_{3,2}')=-J\left(1+e^{i\phi_{3,2}}\right)
\end{equation}
The independence of $F_2$ on $\phi_2'$ and of $F_3$ on $\phi_3'$ means that
\begin{equation}
p_{2,3}(\phi_{2,3}|\vec{\phi}')=p_{2,3}(\phi_{2,3}|\phi'_{3,2})
\end{equation}
so that the total probability distribution separates into independent factors:
\begin{equation}
p(\phi_2,\phi_3)=p_2(\phi_2|\phi_3')p_3(\phi_3|\phi_2')
\end{equation}
Given that the initial probability distribution also factors (as it is simply a uniform distribution), the total probability distribution as given by Eq.~\eqref{eq:ProbabilityPhi} will continue to always be factored, even after many iterations. Furthermore, from the symmetry of the problem, $p_2$ and $p_3$ must necessarily be identical functions, albeit of different arguments. The issue with this is that it is no longer possible to have a probability distribution peaked at $(\phi_2,\phi_3)=(\pm2\pi/3,\mp2\pi/3)$ without introducing additional peaks at $(\mp2\pi/3,\pm2\pi/3)$. We conclude that it is essential to prevent factorization of the probability distribution to avoid such spurious solutions in our scheme. The introduction of diagonal couplings achieves this, with no additional cost. In principle, the consideration of finite-time effects, where the single waveguide probability distribution becomes more complex than that written in Eq.~\eqref{eq:GaussianP} would also prevent factorization.

\section{Numerical procedure}

\paragraph{Generation of the random XY matrices.---}
To benchmark the different optimization schemes, we considered ensembles
of random XY Hamiltonians of different sizes $N$. For each instance, the
independent off-diagonal elements of an initial coupling matrix
$J^{(0)}$ were drawn from a uniform distribution,
\begin{equation}
    J_{ij}^{(0)} \sim \mathcal{U}(-10,10),
    \qquad i<j .
    \label{eq:random_J}
\end{equation}
The matrix was then symmetrized according to
$J_{ji}^{(0)}=J_{ij}^{(0)}$, while the diagonal elements were set to zero,
$J_{ii}^{(0)}=0$.

To keep the characteristic coupling scale comparable between different
matrix sizes, each realization was normalized according to
\begin{equation}
    J_{ij}
    =
    \frac{J_{ij}^{(0)}}
    {\sqrt{
    \frac{1}{N}
    \sum_{k,l}
    \left(J_{kl}^{(0)}\right)^2
    }} .
    \label{eq:J_normalization}
\end{equation}
Consequently,
\begin{equation}
    \frac{1}{N}\sum_{i,j}J_{ij}^{2}=1,
    \label{eq:J_normalization_condition}
\end{equation}
or equivalently $\|J\|_{\mathrm{F}}=\sqrt{N}$, where
$\|\cdot\|_{\mathrm{F}}$ denotes the Frobenius norm.

The pseudorandom numbers were generated using MATLAB's
Mersenne--Twister generator. A base library seed $s_{\mathrm{lib}}$
was first generated as an integer between $1$ and $10^{7}$. The
random-number generator was then reinitialized independently for each
matrix $i$ using the matrix-specific seed
\begin{equation}
    s_i=s_{\mathrm{lib}}+i .
    \label{eq:matrix_seed}
\end{equation}
Both $s_{\mathrm{lib}}$ and $s_i$ were stored together with the matrix,
allowing every realization of the coupling matrix to be reproduced
exactly.

\paragraph{Construction of the $N=50$ reference library.---}
For $N=50$, matrices were generated sequentially and stored in a common
reference library. For each matrix, the reference solution was determined
using the Monte-Carlo plus local-optimization procedure (MC+) introduced
in the main text. We performed
$N_{\mathrm{MC}}=5000$ independent MC+ realizations and retained the
lowest energy found,
\begin{equation}
    E_{\mathrm{ref}}
    =
    \min_{r=1,\ldots,N_{\mathrm{MC}}}
    E_{\mathrm{MC}}^{(r)} ,
    \label{eq:E_ref}
\end{equation}
together with its corresponding phase configuration. The use of a large
number of independent realizations provides a high-confidence
best-known reference solution for each random XY instance.

The same coupling matrix was subsequently solved using
$N_{\mathrm{GD}}=1000$ independent Gain-D trajectories. Gain-D was
treated strictly as a benchmark: its results were not used to modify
$E_{\mathrm{ref}}$. Instead, every Gain-D trajectory was compared with
the fixed MC+ reference solution.

For any solver, a realization producing a final energy $E$ was classified
as successful when
\begin{equation}
    \frac{|E-E_{\mathrm{ref}}|}
    {\max\!\left(|E_{\mathrm{ref}}|,
    \epsilon_{\mathrm{mach}}\right)}
    \leq \epsilon_{\mathrm{tol}},
    \label{eq:success_condition}
\end{equation}
where we used
\begin{equation}
    \epsilon_{\mathrm{tol}}=10^{-8},
\end{equation}
corresponding to a relative energy accuracy of $10^{-6}\%$.

For a given matrix $i$ and solver $s$, the single-realization success
probability was estimated directly from the fraction of successful
trajectories,
\begin{equation}
    p_i^{(s)}
    =
    \frac{N_{\mathrm{succ},i}^{(s)}}
    {N_{\mathrm{run},i}^{(s)}} .
    \label{eq:single_success_probability}
\end{equation}
In particular, the MC+ success probability was evaluated using the same
$5000$ realizations employed to determine $E_{\mathrm{ref}}$.

To ensure that the reference library contained a sufficient number of
difficult optimization instances, a matrix was classified as
``hard'' when
\begin{equation}
    p_i^{(\mathrm{MC})}<0.1 .
\end{equation}
Random matrices were continuously generated and appended to the library
until at least $200$ such hard instances had been accumulated. Importantly,
all generated matrices were retained, not only the hard instances. The
resulting $N=50$ reference library contained $12,205$ random XY
Hamiltonians.

For each matrix, the coupling matrix, random seed, MC+ reference energy
and phase configuration, individual MC+ and Gain-D trajectory energies,
success probabilities, and relevant numerical parameters were stored.
The waveguide (WG) calculations were subsequently performed using these
stored matrices and the same fixed reference energies and success
criterion of Eq.~\eqref{eq:success_condition}.

\paragraph{Success probability after repeated realizations.---}
The quantity $p_i^{(s)}$ characterizes the probability that a single
realization of solver $s$ successfully solves matrix $i$. Assuming
independent realizations, the probability that at least one of $n$
realizations is successful is therefore
\begin{equation}
    q_i^{(s)}(n)
    =
    1-\left(1-p_i^{(s)}\right)^n .
    \label{eq:cumulative_success}
\end{equation}
To characterize the performance of a solver over an ensemble containing
$N_{\mathrm{mat}}$ matrices, we averaged this quantity with equal weight
over all instances,
\begin{equation}
    P^{(s)}(n)
    =
    \frac{1}{N_{\mathrm{mat}}}
    \sum_{i=1}^{N_{\mathrm{mat}}}
    q_i^{(s)}(n).
    \label{eq:ensemble_success}
\end{equation}
We stress that Eq.~\eqref{eq:ensemble_success} averages the cumulative
success probability of each matrix individually. It is therefore not
equivalent to first averaging the single-realization probabilities
$p_i^{(s)}$ and subsequently applying
$1-(1-p)^n$. This distinction preserves the matrix-to-matrix variation
in optimization difficulty.

\begin{figure}[t!]
\includegraphics[width=0.7\textwidth]{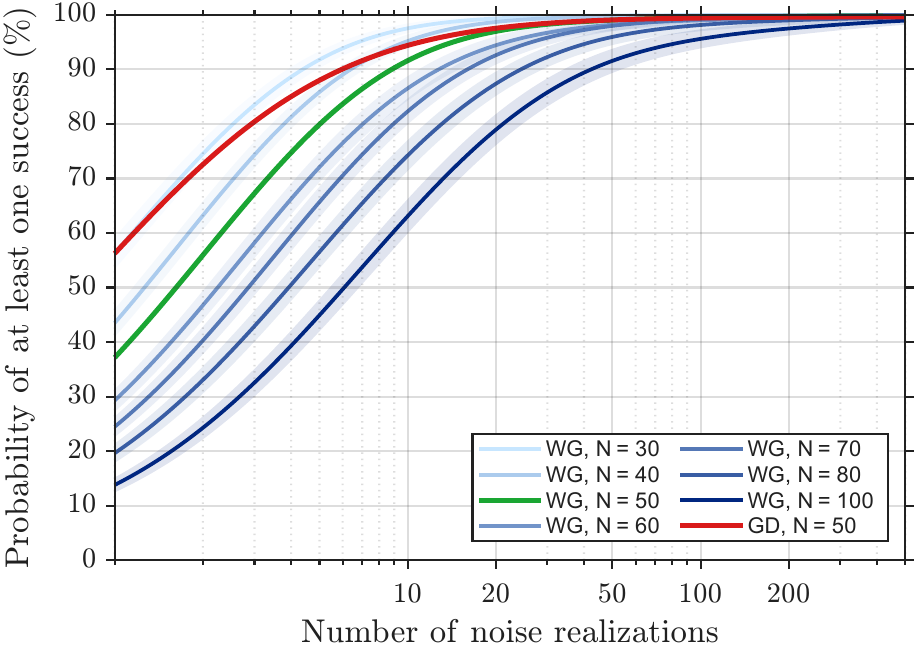}
\caption{\textbf{Probability of obtaining at least one successful realization as a function of the number of independent noise realizations.} For each matrix, the probability after $n$ independent runs is calculated as $1-(1-p_i)^n$, where $p_i$ is the corresponding single-run success probability, and the result is then averaged over the matrix ensemble. Blue curves show the WG results for different matrix sizes $N$, with $N=50$ highlighted in green. The red curve shows Gain-D for $N=50$ using the same matrix ensemble. Shaded regions indicate the $99\%$ confidence interval of the matrix-averaged probability. The zoomed high-success region for $20\leq n\leq500$ and $0.95\leq P\leq1$ is shown in Fig.~5 of the main text.}
\label{fig:Scalability_full}
\end{figure}

For the direct comparison between WG and Gain-D at $N=50$, both methods
were evaluated over exactly the same complete ensemble of $12\,205$
matrices. Consequently, differences between the WG and Gain-D curves
cannot arise from different sampled matrix populations. For the WG
matrix-size scaling shown in Fig.~\ref{fig:Scalability} and Fig.~\ref{fig:Scalability_full},
the same analysis was independently repeated for the available matrix
sizes from $N=30$ to $N=100$.

The shaded regions in Fig.~\ref{fig:Scalability} represent
$99\%$ confidence intervals for the ensemble mean. At each $n$, we first
calculated the standard deviation $\sigma_q^{(s)}(n)$ of
$q_i^{(s)}(n)$ across the matrix ensemble. The standard error of the mean
was then
\begin{equation}
    \mathrm{SE}^{(s)}(n)
    =
    \frac{\sigma_q^{(s)}(n)}
    {\sqrt{N_{\mathrm{mat}}}} ,
    \label{eq:SE_success}
\end{equation}
and the plotted $99\%$ confidence interval was taken as
\begin{equation}
    P^{(s)}(n)
    \pm
    2.576\,\mathrm{SE}^{(s)}(n),
    \label{eq:CI_success}
\end{equation}
with the interval restricted to the physical range $[0,1]$.

\paragraph{Waveguide simulation parameters and optimization.---}

The WG dynamics were integrated directly using the equations of motion introduced in the main text. For the numerical results reported here, the normalized coupling matrix was multiplied by an overall coupling strength
\begin{equation}
    g = 0.020 .
\end{equation}
The pulsed protocol had a period
\begin{equation}
    T = 7\pi ,
\end{equation}
with the gain active during the first fraction
$f_{\mathrm{pump}}=0.75$ of each period. The feedback was activated only after the pump was switched off and remained active for the remainder of the cycle. The propagation delay was therefore chosen as
\begin{equation}
    \tau=(1-f_{\mathrm{pump}})T=\frac{T}{4},
\end{equation}
and the feedback window was set equal to this delay,
$R_{\mathrm{window}}=\tau$. The maximum feedback amplitude was $R_0=1$.

The gain parameter was set to $P_0=15$. In the numerical implementation, the gain amplitude was slowly increased over the full simulation from $P_{\mathrm{base}}=0.8$ toward $P_0$, while retaining the square-wave on--off modulation within each period. This slow ramp reduces sensitivity to the initial transient while preserving the repeated condensation and feedback protocol. The feedback was strictly gated such that the pump and coupling stages did not overlap.

\subsection{Optimization of the waveguide parameters}

The parameters of the waveguide (WG) solver were optimized prior to the
large-scale benchmark. The aim of this preliminary optimization was to identify
a single set of WG parameters that could subsequently be kept fixed for all
matrices, rather than adapting the dynamics independently to each optimization
instance. The optimization was performed for matrices of size $N=50$, which is
the principal system size considered in the benchmark.

A fixed training set of $10$ matrices was randomly selected from the $N=50$
matrix library using a fixed random seed. The same set of matrices was used
throughout the parameter search. For every candidate parameter set, the WG
dynamics was simulated using $300$ independent trajectories for each training
matrix. The integration time step was fixed to
\[
\Delta t = 0.01,
\]
and the feedback amplitude was fixed to
\[
R_0 = 1.
\]
The remaining WG parameters were treated as hyperparameters and optimized
numerically.

The optimization was carried out using a sequential coordinate-search
procedure. Starting from the central value of each parameter range, one
parameter was varied while all other parameters were kept fixed at their
current optimal values. The value yielding the highest score was retained
before proceeding to the next parameter. The parameters were optimized in the
order
\[
\delta_{\mathrm{diag}}
\rightarrow
f_{\mathrm p}
\rightarrow
T
\rightarrow
P_0
\rightarrow
g
\rightarrow
\sigma ,
\]
where $\delta_{\mathrm{diag}}$ denotes an optional diagonal offset applied to
the coupling matrix, $f_{\mathrm p}$ is the pump fraction, $T$ is the
characteristic temporal period of the WG dynamics, $P_0$ is the pump
amplitude, $g$ sets the overall coupling scale, and $\sigma$ is the noise
amplitude.

The search ranges used in the optimization were
\[
\delta_{\mathrm{diag}}
\in
\{0,0.5,1,\ldots,5\},
\]
\[
f_{\mathrm p}
\in
\{0.73,0.74,0.75\},
\qquad
T
\in
\{5\pi,7\pi,10\pi\},
\]
\[
P_0
\in
\{5,10,15\},
\qquad
g
\in
\{0.016,0.018,0.020\},
\]
and
\[
\sigma
\in
\{10^{-5},10^{-4},10^{-3}\}.
\]

For a given candidate parameter set, the final XY energy of every WG
trajectory was compared with the reference energy associated with the
corresponding matrix in the library. Defining the relative energy error as
\[
\varepsilon_E
=
\frac{|E-E_{\mathrm{ref}}|}
{|E_{\mathrm{ref}}|},
\]
a trajectory was considered successful during the parameter-optimization
stage when
\[
\varepsilon_E \leq \varepsilon_{\mathrm{opt}},
\]
with $\varepsilon_{\mathrm{opt}}=10^{-5}$. For each matrix, the candidate
parameter set was considered to solve the instance if at least one of its WG
trajectories satisfied this condition.

The score used for the coordinate search was
\[
\mathcal{S}
=
1000\,N_{\mathrm{solved}}
+
P_{\mathrm{single}}
-
10\,N_{\mathrm{unsolved}},
\]
where $N_{\mathrm{solved}}$ and $N_{\mathrm{unsolved}}$ are respectively the
numbers of training matrices for which at least one successful trajectory was
or was not obtained, and $P_{\mathrm{single}}$ is the fraction of successful
trajectories over the complete training ensemble. The large weight assigned
to $N_{\mathrm{solved}}$ makes the optimization primarily favor parameter sets
that solve the largest possible number of distinct matrices. The
single-trajectory success probability then provides an additional criterion
for discriminating between parameter sets having comparable matrix coverage.

Some temporal quantities were not independently optimized but were derived
from the pump fraction and the period $T$. In particular,
\[
\tau=(1-f_{\mathrm p})T,
\]
and the coupling window was chosen as
\[
R_{\mathrm{window}}=\tau.
\]
The total propagation time was fixed to
\[
t_{\max}=200T+\frac{T}{2}.
\]
Consequently, changes in $T$ and $f_{\mathrm p}$ simultaneously modify the
absolute duration of the dynamical stages while preserving the prescribed
relation between the pump and coupling windows.

We additionally tested a diagonal shift of the form
\[
J \rightarrow J+\delta_{\mathrm{diag}} I.
\]
Since such a term shifts the energy of all XY phase configurations by a constant, the corresponding reference energy was corrected consistently when evaluating the optimization score. The optimization selected

\[
\delta_{\mathrm{diag}}=0,
\]
so that no artificial diagonal shift was retained in the final WG calculations.

The parameter set retained for the production simulations was
\[
P_0=15,
\qquad
T=7\pi,
\qquad
f_{\mathrm p}=0.75,
\]
\[
g=0.020,
\qquad
\sigma=10^{-5},
\qquad
\delta_{\mathrm{diag}}=0.
\]
These values correspond to
\[
\tau=0.25T,
\qquad
R_{\mathrm{window}}=\tau,
\]
with
\[
R_0=1,
\qquad
\Delta t=0.01,
\qquad
t_{\max}=200T+\frac{T}{2}.
\]

The final benchmark calculations were then performed with these parameters kept fixed and with $1000$ independent WG trajectories per matrix.

We emphasize that the tolerance $\varepsilon_{\mathrm{opt}}=10^{-5}$ was used only during the hyperparameter search. The success probabilities reported in the final benchmark were recomputed using the stricter common criterion
\[
\frac{|E-E_{\mathrm{ref}}|}
{\max(|E_{\mathrm{ref}}|,\epsilon_{\mathrm{mach}})}
\leq 10^{-8},
\]
so that the comparison between WG, Gain-D, and the reference optimization procedure is performed using the same definition of success.

\subsection{Success probability under repeated realizations}

Figure~S1 shows the probability of obtaining at least one successful trajectory as a function of the number of independent noise realizations. For a matrix with single-run success probability $p_i$, the probability of
obtaining at least one successful realization after $n$ independent runs is
\[
P_i(n)=1-(1-p_i)^n.
\]
The curves shown in Fig.~6 are obtained by averaging this quantity over the corresponding ensemble of matrices, with the shaded regions indicating the $99\%$ confidence interval of the matrix-averaged probability.

For the WG solver, the success probability decreases systematically with increasing matrix size $N$, as expected from the increasing complexity of the optimization landscape. Nevertheless, repeated independent realizations rapidly compensate for the reduced single-shot probability: for all matrix sizes considered here, the probability of obtaining at least one successful trajectory approaches unity as the number of realizations is increased. The $N=50$ WG result is highlighted in green and compared directly with Gain-D for the same $N=50$ matrix ensemble, shown in red.

The full range displayed here emphasizes the different initial single-shot success probabilities and the convergence toward high total success probability with repeated runs. Figure~5 of the main text shows an enlarged view of the high-success regime, restricted to $20\leq n\leq500$ and $0.95\leq P\leq1$, in order to resolve the differences between the methods once their cumulative success probabilities are already close to unity.

\section{Dynamical evolution toward the XY ground state}

Figure 2 illustrates the temporal evolution of a representative successful realization of the waveguide spin simulator. The alternating pump and feedback stages progressively establish the condensate amplitudes and reorganize their relative phases. During the first few cycles, the phases remain strongly fluctuating while the field builds up. Repeated feedback then rapidly reduces the XY energy, after which the phase configuration stabilizes close to the ground-state solution.

\begin{figure}[t!]
\includegraphics[width=0.8\textwidth]{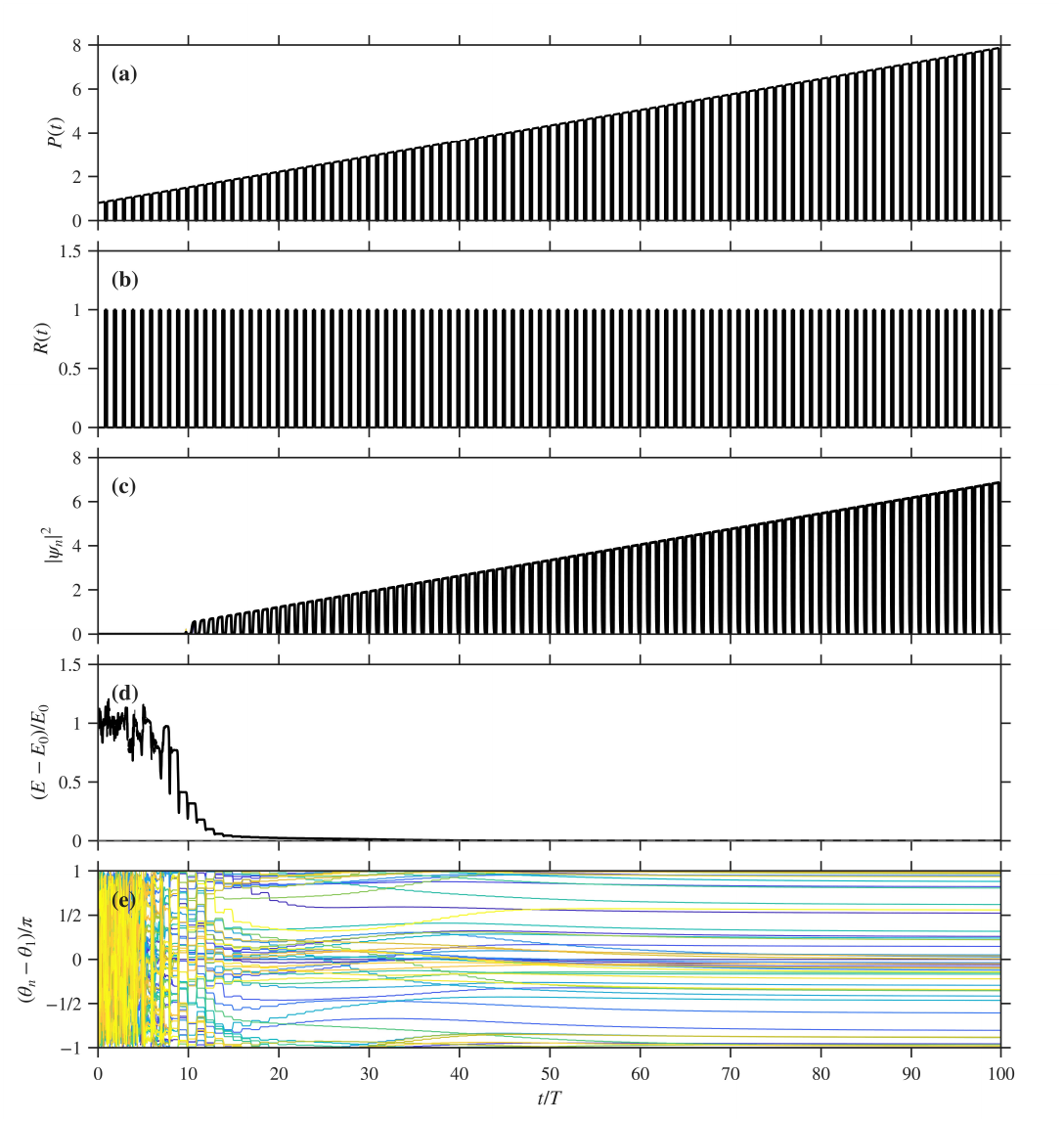}
\caption{Dynamical evolution of a representative successful 
waveguide-spin-simulator realization. 
(a) Time-dependent pump \(P(t)\), whose amplitude increases gradually during the simulation. 
(b) Feedback coefficient \(R(t)\), activated periodically during the coupling stage of each cycle. 
(c) Intensities \(\lvert\psi_n\rvert^2\) of the \(N=50\) waveguide modes. 
(d) Relative XY energy \((E-E_0)/E_0\), showing rapid relaxation toward the ground-state energy \(E_0\). 
(e) Relative phases \((\theta_n-\theta_1)/\pi\). 
The initially fluctuating phases progressively organize under repeated feedback and settle into a stable configuration as the XY energy approaches its minimum.}
\label{fig:Time_dynamics}
\end{figure}


\end{document}